\documentclass[fleqn,twoside]{article}
\usepackage{espcrc2}

\usepackage{graphicx}
\usepackage{hhline}
\usepackage{epsfig}
\usepackage{amsmath}
\let\cl=\centerline
\def\figbox#1;#2;{\parbox{#2cm}{%
\vglue3mm\epsfig{file=#1.eps,width=#2cm}\vglue3mm}}
\def\figboxc#1;#2;{\cl{\figbox #1;#2;}}
\usepackage[figuresright]{rotating}

\newcommand{\piz}{\ensuremath{\pi^0\,}}

\newcommand{\etatrepi}{\ensuremath{\eta\rightarrow3\pi}}

\title{\bf{ Measurement of the slope parameter $\boldsymbol{\alpha}$ \\
    \bf  for the $\boldsymbol{\eta\rightarrow 3\piz}$ decay at KLOE}}
\author{KLOE collaboration:
F.~Ambrosino,
A.~Antonelli, 
M.~Antonelli, 
F.~Archilli,
C.~Bacci,
P.~Beltrame,
G.~Bencivenni, 
S.~Bertolucci, 
C.~Bini, 
C.~Bloise, 
S.~Bocchetta, 
V.~Bocci,
F.~Bossi,
P.~Branchini,
R.~Caloi,
P.~Campana, 
G.~Capon, 
T.~Capussela,
F.~Ceradini,
S.~Chi,
G.~Chiefari, 
P.~Ciambrone,
E.~De~Lucia,
A.~De~Santis, 
P.~De~Simone, 
G.~De~Zorzi,
A.~Denig,
A.~Di~Domenico,
C.~Di~Donato,
S.~Di~Falco,
B.~Di~Micco,
A.~Doria,
M.~Dreucci,
G.~Felici, 
A.~Ferrari,
M.~L.~Ferrer, 
G.~Finocchiaro,
S.~Fiore,
C.~Forti,       
P.~Franzini,
C.~Gatti,      
P.~Gauzzi,
S.~Giovannella,
E.~Gorini, 
E.~Graziani,
M.~Incagli,
W.~Kluge,
V.~Kulikov,
F.~Lacava, 
G.~Lanfranchi, 
J.~Lee-Franzini,
D.~Leone,
M.~Martini,
P.~Massarotti,
W.~Mei,
S.~Meola,
S.~Miscetti, 
M.~Moulson,
S.~M\"uller,
F.~Murtas, 
M.~Napolitano,
F.~Nguyen,
M.~Palutan,          
E.~Pasqualucci,
A.~Passeri,  
V.~Patera,
F.~Perfetto,
M.~Primavera,
P.~Santangelo,
G.~Saracino,
B.~Sciascia,
A.~Sciubba,
F.~Scuri, 
I.~Sfiligoi,     
T.~Spadaro,
M.~Testa,
L.~Tortora, 
P.~Valente,
B.~Valeriani,
G.~Venanzoni,
R.~Versaci,
G.~Xu.
}
\begin{document}
\begin{abstract}
\begin{center}
\noindent
    {\bf Abstract}\\
\end{center}
We report a preliminary measurement of the slope parameter $\alpha$
for the $\eta\rightarrow 3\piz$ decay carried out with KLOE at
DA$\Phi$NE; where $\alpha$ is the parameter describing the energy
dependence of the square of the matrix element for this decay. By
fitting the event density in the Dalitz plot with a collected
statistic of 420 pb$^{-1}$ we determine $\alpha = -0.027 \pm
0.004\,(stat) \;^{ +0.004}_{-0.006}\,(syst)$. This result is
consistent with current chiral perturbation theory calculations within
the unitary approach. 
\vspace{1pc}
\end{abstract}
% typeset front matter (including abstract)
\maketitle
\section{Introduction}
With a branching ratio of $32$\% the $\eta\rightarrow 3\piz$ decay is
a major decay mode of the $\eta$ despite the fact that is a G--parity
forbidden transition. Neglecting a small electromagnetic transition
(Sutherland's theorem \cite{Sutherland}) this decay is due almost
exclusively to the isospin breaking part of QCD:
\begin{equation}
{\cal L}_{\not\, I} = -\frac{1}{2}\left(m_{u}-m_{d}\right)
\left(\bar{u}u-d\bar{d}\right) 
\end{equation}
\noindent
and provides a nice way to determine the up-down quark mass difference.\\
Many theoretical issues \cite{BijGa02}, \cite{Hol02} refer to the
study of the charged and neutral decay mode of \etatrepi, and in
particular to the Dalitz plot parameters of these decays.\\
The Dalitz plot distribution of the $\eta\rightarrow 3\piz$ decay is
conventionally described in terms of one kinematical variable:
\begin{equation}
z = \frac{2}{3} \sum_{i=1}^{3} \left (\frac{3E_{i} -
m_{\eta}}{m_{\eta} - 3m_{\piz}} \right )^{2} =
\frac{\rho^{2}}{\rho_{MAX}^{2}} 
\label{eq:zeta}
\end{equation}
\noindent
where $E_{i}$ denotes the energy of the i-th pion in the $\eta$ rest
frame and $\rho$ is the distance from a point on the Dalitz plot to
its center. $\rho_{MAX}$ is the maximum value of $\rho$.
For the decays into three identical particles, it is possible to use a
symmetrical Dalitz plot where the event density is described by a
single quadratic slope parameter $\alpha$, which represents the
difference from pure phase space:
\begin{equation}
\vert A_{\eta\to 3\piz}\left(z\right) \vert^{2} \sim 1 + 2\alpha z.
\end{equation}
\noindent
The lowest order  predictions of Chiral Pertubation Theory quote a
zero value for $\alpha$. The event density in the Dalitz plot must be
uniform. A non zero value is instead expected by dispersive
calculation \cite{BaKaWy96} where the effect of $\pi-\pi$ rescattering
is included. Lately a theoretical result \cite{Borasoy}, obtained in
the chiral unitary approach based on the Bethe--Salpter equation,
provide for $\alpha$ the value $-0.031$. 
There are three previous experimental determination of $\alpha$: the
GAMS 2000 group \cite{GAMS} quoted $\alpha = -0.022 \pm 0.023$ based
on $5 \times 10^{4}$ events; the Crystall Barrel Collaboration
obtained $\alpha = -0.052 \pm 0.020$ from a sample of  $10 \times
10^{4}$ events and finally the Crystal Ball \cite{CrystalBall} result,
$\alpha = -0.031 \pm 0.004$ based on $10^{6}$ events.
This was the first statistically significant measurement of $\alpha$
which agree very well with the most recent theoretical result. 
In this paper we report on a new precise measurement of the Dalitz
plot parameter for the $\eta\rightarrow 3\piz$ decay.\\
\section{DA$\Phi$NE and KLOE}
\noindent 
The DA$\Phi$NE e$^+$e$^-$ collider operates at a total energy W = 1020
MeV, the mass of the $\phi$(1020)--meson.
Approximately $3\times10^6$ $\phi$--mesons are produced for each
pb$^{-1}$ of collected luminosity.
Since 2001, KLOE has integrated a total luminosity of about 2.5
fb$^{-1}$. 
Results presented in this paper are based on data collected on
$2001-2002$ only and correspond to about 420~pb$^{-1}$.
The KLOE detector consists of a large cylindrical drift chamber, DC,
surrounded by a lead/scintillating-fiber electromagnetic calorimeter,
EMC.  
The drift chamber \cite{dc}, is 4~m in diameter and 3.3~m long.
The momentum resolution is $\sigma(p_{T})/p_{T} \sim 0.4\%$. 
Two track vertices are reconstructed with a spatial resolution of
$\sim$ 3 mm. 
The calorimeter \cite{emc}, composed of a barrel and two
endcaps,covers 98\% of the solid angle.
Energy and time resolution are $\sigma(E)/E = 5.7\%/\sqrt{E[{\rm
    GeV}]}$ and $\sigma(t) = 57 \,{\rm ps}/ \sqrt{E[{\rm GeV}]} \oplus
100 \, {\rm ps}$.
A superconducting coil around the detector provides a 0.52~T magnetic
field.
The KLOE trigger \cite{trg}, uses calorimeter and drift chamber
information.
For the present analysis only the electromagnetic calorimeter (EMC) 
signals have been used. Two local energy deposits above threshold, 
$E_{\rm th}>50$ MeV for the barrel and $E_{\rm th}>150$ MeV for the 
endcaps, are required. 
\section{ Dalitz plot of  $\eta\rightarrow\piz\piz\piz$ decay}
\begin{figure}[!h] 
  \begin{center}
    \epsfig{figure=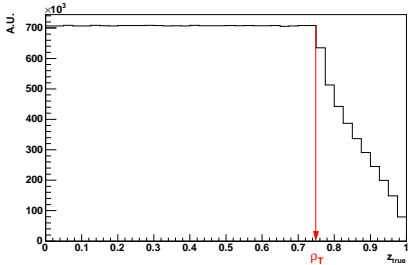,width=.8\linewidth}
    \caption{MC $z$ distribution according to pure phase space.} 
    \label{fig:DalitzPlot} 
  \end{center}
\end{figure}

\begin{figure}[!h] 
  \begin{center}
    \epsfig{figure=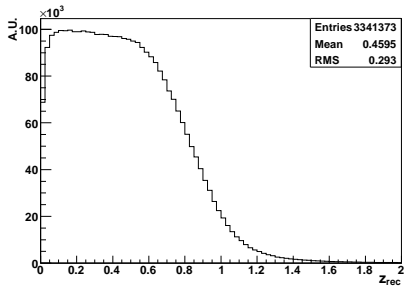,width=.8\linewidth}
    \caption{MC $z$ distribution after selection and reconstruction.} 
    \label{fig:DalitzPlot1} 
  \end{center}
\end{figure}

At KLOE the $\eta$ meson is produced in the process
$\phi\rightarrow\eta\gamma$ where the recoil photon ($E_{\gamma} = 363
\,\textrm{MeV})$ is monochromatic and easily selected.
Thus to select the final state we require to have seven prompt
clusters in the event. 
After applying a kinematic fit, which requires the energy-momentum
conservation, we look for the best pairing of photons into $\piz$ by
constructing a pseudo--$\chi^{2}$ variable for each of the $15$
possible pairs:\\
\begin{equation}
\chi^{2}_{j} = \sum_{i=1}^{3} \left( \frac{m_{j,\pi^{0}_{i}} - M_{\pi^{0}}}
    {\sigma_{m_{\pi^{0}}}}\right)^{2} \qquad j = 1,2,\ldots,15.
\end{equation}
\noindent
where 
\begin{itemize}
\item[-] $m_{j,\pi^{0}_{i}}$ is the invariant mass of the
$i^{th} \pi^{0}$ , in correspondence of the $j^{th}$ combination; 
\item[-] $M_{\pi^{0}}$ is the $\pi^{0}$ mass, ($M_{\pi^{0}}$
  = 134.98 MeV/c$^{2}$ \cite{PDG}); 
\item[-] $\sigma_{m_{\pi^{0}}}$ is the resolution on $m_{\pi^{0}}$. 
\end{itemize} 
The chosen pair is that one which minimize the $\chi^{2}$.
A second kinematic fit constraining the $\piz$ mass has been
performed, improving the resolution on $z$ by a factor two.
The Monte Carlo, MC, $z$ distribution at generation (pure phase
space), see fig.\ref{fig:DalitzPlot} and after reconstruction, see
fig.\ref{fig:DalitzPlot1}, shows that resolution effects are not
negligible for this analysis. 
Three samples with different efficiency, $\varepsilon$ and purity on
pairing, $P$: 
\begin{align*}
&\textrm{Low Purity}    & &P = 75.4 \%  & &\varepsilon = 30.3 \%\\
&\textrm{Medium Purity} & &P = 92.0 \%  & &\varepsilon = 13.6 \%\\
&\textrm{High Purity}   & &P = 97.6 \%  & &\varepsilon = 4.3  \%\\
\end{align*}
have been analyzed by cutting on the difference of the two lowest
value of $\chi^{2}$.
\section{Measurement of the slope parameter $\alpha$}
\begin{figure}[htb] 
  \begin{center}
      \epsfig{figure=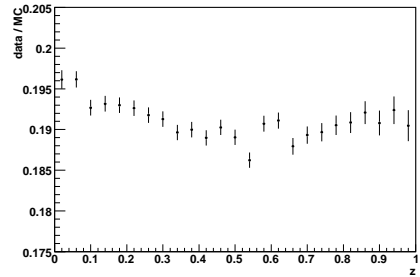,width=.8\linewidth}
  \end{center} 
  \caption{Data--Monte Carlo ratio of the $z$ distribution. The MC
    distribution is pure phase space.}
  \label{fig:linearita} 
\end{figure}

\begin{figure}[htb] 
  \begin{center}
      \epsfig{figure=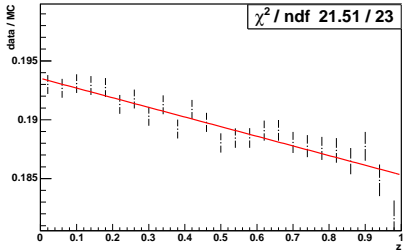,width=.8\linewidth}
  \end{center} 
  \caption{Data--Monte Carlo ratio of the $z$ distribution after the
    kinematic fit with $\eta$ mass constraint.}
  \label{fig:linearitaII} 
\end{figure}

In order to estimate $\alpha$ an unbinned likelihood function is built
by convoluting the event density with the resolution function and
correcting for the probability of wrong photon pairing in $\piz$'s. 
Using a sample with high purity and fitting in the range $(0-1)$, we
found the preliminary result\cite{tiziana}:
\begin{equation}
  \alpha = -0.014 \pm 0.004\,(stat) \;\pm 0.005\,(syst),
\end{equation}
where the systematics has been evaluated by varying the analysis cuts,
the fit range and measuring the maximum observed variation of $\alpha$
with respect to the samples with different purity.\\
We observed a relevant dependence of $\alpha$ on the fitting
range. Moreover, the low purity sample shows two different slopes (see 
fig.\ref{fig:linearita}) in the Data--MC ratio of the $z$ distribution.
With a dedicated simulation we have realized that this is essentially
due to a different value of invariant mass of three pions system
$(\piz\piz\piz)$ in the Monte Carlo generator, $M_{\eta} = 547.30$
MeV/c$^{2}$, with respect to the one recently measured \cite{Biagio}
by our experiment:
\begin{equation}
M_{\eta} =  547.822 \pm 0.005_{stat} \pm 0.069_{syst} \,\textrm{MeV/c$^{2}$}.
\label{eq:Biagio}
\end{equation}
As a consequence, the accessible phase space on data is larger than
the one on Monte Carlo simulation.
A new evaluation of $\alpha$ has been obtained after a kinematic fit
with an additional constraint on the $\eta$ mass, see Table
\ref{tab:resfit_3}.
\begin{table}[!hab]
  \begin{center}
    \begin{tabular}{||c||c||c||c||}
      \hhline{|t:=:t:=:t:=:t:=:t|}
      \multicolumn{1}{||c||}{\small{\bf Range}} &  
      \multicolumn{1}{c||}{\small{\bf Low Pur. }} &
      \multicolumn{1}{c||}{\small{\bf Med. Pur.}}&
      \multicolumn{1}{c||}{\small{\bf High Pur.}}\\ 
      & $\alpha (\cdot 10^{-3})$ & $\alpha (\cdot 10^{-3})$ & $\alpha
      (\cdot 10^{-3})$ \\
      \hhline{|:=::=::=::=:|}
      (0,1)   &$-30\pm 2$ &$-31\pm 3$  &$-26\pm 4$\\
      (0,0.8) &$-26\pm 2$ &$-28\pm 3$  &$-22\pm 5$\\
      (0,0.7) &$-26\pm 3$ &$-27\pm 4$  &$-23\pm 5$\\
      (0,0.6) &$-30\pm 4$ &$-31\pm 4$  &$-20\pm 6$\\
      \hhline{|b:=:b:=:b:=:b:=:b|}
    \end{tabular}
  \end{center}
  \caption{Fitted results for the slope parameter $\alpha$ for the
    kinematic fit with the $\eta$ mass constrained at $M_{\eta} =
    547.822$ MeV/c$^{2}$.}
  \label{tab:resfit_3}
\end{table}\\
\noindent
In this way, good stability with respect to the range of fit is
observed, and the linearity of the Data--MC ratio of the z
distribution has been recovered, (see fig.\ref{fig:linearitaII}).
In order to give the final result, the phase space area where the $z$
distribution is flat, is used as fit region (z $\in [0,0.7]$). 
The systematic uncertainties on $\alpha$ are summarized in Table
\ref{tab:syst}. We have first taken into account the correction for
the Data--MC discrepancy in the photon energy resolution, RES, the
effect of the fit range. We have then determined the $\alpha$
variation with respect to other purity samples and the error due to
the combinatorial background evaluation.
We have also tested the $\alpha$--dependence on the chosen value of
the $M_{\eta}$ used in the fit constraint.
\begin{table}[htb]
  \begin{center}
    \begin{tabular}{||l||c||c||c||}
      \hhline{|t:=:t:=:t:=:t:=:t|}
      \multicolumn{1}{||c||}{\small{\bf }} &  
      \multicolumn{1}{c||}{\small{\bf Low Pur. }} &
      \multicolumn{1}{c||}{\small{\bf Med. Pur.}}&
      \multicolumn{1}{c||}{\small{\bf High Pur.}}\\ 
      & $\sigma^{syst}_{\alpha} (\cdot 10^{-3})$ & $\sigma^{syst}_{\alpha} (\cdot 10^{-3})$ & $\sigma^{syst}{\alpha}
      (\cdot 10^{-3})$ \\
      \hhline{|:=::=::=::=:|}
      RES        &$-9$       &$-4$        &$-3$\\
      Range      &$-4$       &$-4$        &$-3 +3$\\
      Purity     &$-1 +3$&$+4$        &$-4 $\\
      BKG        &$0.$           &$0.$            &$-1 +1$ \\
      $M_{\eta}$ &$-1$       &$-2$        &$-5$ \\
      \hhline{|t:=:t:=:t:=:t:=:t|}
      {\bf Total } &$-10 \,+3$ &$-6 \,+4$ &$-8 \,+3$ \\
      \hhline{|b:=:b:=:b:=:b:=:b|}
    \end{tabular}
  \end{center}
  \caption{Summary table of systematic uncertainties. The total
    systematic uncertainty is obtained by adding in quadrature each
    contribution.}
  \label{tab:syst}
\end{table}    
\section{Results}
\noindent
We quote as preliminary result for the slope parameter $\alpha$ the
one obtained with the Medium Purity sample ,(about 650000
$\eta\rightarrow3\pi^{0}$ decays).  
The result including the statistical uncertainty from the fit and the
evaluated systematic error is:
\begin{equation}
  \alpha = -0.027 \pm 0.004\,(stat) \;^{+0.004}_{-0.006}\,(syst)
\end{equation}
with: $\chi^{2}/ndf = 13.72/17$.
The result is within errors compatible with the result from Crystall
Ball \cite{CrystalBall} based on $10^{6}$ events of:
\begin{equation}
  \alpha = -0.031 \pm 0.004.
\end{equation}
The our  measurement of $\alpha$ also agrees with the
calculations from the chiral unitary approach \cite{Borasoy}.


\begin{thebibliography}{9}
\bibitem{Sutherland}
  D.G.~Sutherland, Phys.Lett. {\bf 23}, 384 (1966). 
%\bibitem{tesiPersson}
%  J.~Bijnens and F.~Persson, [hep-ph/0106130]. 
\bibitem{BijGa02}
  J.~Bijnens and J.~Gasser, Physica Scripta {\bf T99}, 34 - 44 (2002).
\bibitem{Hol02} 
  B.~Holstein, Physica Scripta {\bf T99}, 55 - 67, (2002).
\bibitem{BaKaWy96}
  R.~Baur, J.~Kambor and D.~Wyler Nucl. Phys. {\bf B460},
  127 (1966).
\bibitem{Borasoy}
  B.~Borasoy, R.~Nissler, [hep-ph/0510384 v2]. 
\bibitem{GAMS}
  GAMS 2000 Collaboration, D. Alde {\em et al.}, Phys. C {\bf 25}, (1984) 225.
\bibitem{CrystalBarrel}
  Crystal Barrel Collaboration, A. Abele {\em et al.}, Phys.Lett. B {\bf 417}, (1998) 193.
\bibitem{CrystalBall}
  Crystal Ball Collaboration, Phys. Rev. Lett. {\bf 87}, (2001) 19. 
\bibitem{dc}
  M.~Adinolfi {\em et al.}, \emph{The tracking detector of the KLOE experiment}, Nucl. Instrum. Meth {\bf A 488} {2002} 51.
\bibitem{emc} 
  M.~Adinolfi {\em et al.}, \emph{The KLOE electomagnetic calorimeter},
  Nucl. Instrum. Meth {\bf A 482} {2002} 364
\bibitem{trg} 
  M.~Adinolfi {\em et al.}, \emph{The trigger system of the KLOE experiment},
  Nucl. Instrum. Meth {\bf A 492} {2002} 134
\bibitem{PDG}  W.-M. Yao {\em et al.} Journal of Physics  {\bf G 33, 1} {2006}
\bibitem{Biagio}
  B. Di Micco [KLOE Collaboration],
  {\it Acta Phys. Slov.} {\bf 56} (2005) 403-409.
\bibitem{tiziana}
  T. Capussela [KLOE Collaboration],
  {\it Acta Phys. Slov.} {\bf 56} (2005) 341-344.
\end{thebibliography}
\end{document}